\documentclass[amsmath,amssymb,prd,preprintnumbers,showpacs]{revtex4}

\usepackage{graphicx}

\newcommand{\beq}[1]{\begin{equation}\label{#1}}
\newcommand{\eeq}{\end{equation}}
\newcommand{\bea}[1]{\begin{eqnarray}\label{#1}}
\newcommand{\eea}{\end{eqnarray}}

\def\ktr{\tilde{\kappa}_{\rm tr}}
\def\etal{{\it et al.}}

\def\lsim{\mathrel{\rlap{\lower3pt\hbox{$\sim$}}
    \raise2pt\hbox{$<$}}}
\def\gsim{\mathrel{\rlap{\lower3pt\hbox{$\sim$}}
    \raise2pt\hbox{$>$}}}

\newcommand{\rf}[1]{(\ref{#1})}

\begin{document}

\title{Violations of Einstein's Relativity: Motivations, Theory, and Phenomenology}

\author{Ralf Lehnert}
\affiliation{Instituto de Ciencias Nucleares,
Universidad Nacional Aut\'onoma de M\'exico,
A.~Postal 70-543, 04510 M\'exico D.F., Mexico}

\date{February 16, 2011}

\begin{abstract} 
One of the most difficult questions in present-day physics concerns a fundamental theory of space, time, and matter that incorporates a consistent quantum description of gravity. There are various theoretical approaches to such a quantum-gravity theory. Nevertheless, experimental progress is hampered in this research field because many models predict deviations from established physics that are suppressed by some power of the Planck scale, which currently appears to be immeasurably small.

However, tests of relativity theory provide one promising avenue to overcome this phenomenological obstacle: many models for underlying physics can accommodate a small breakdown of Lorentz symmetry, and numerous feasible Lorentz-symmetry tests have Planck reach. Such mild violations of Einstein's relativity have therefore become a focus of recent research efforts.

This presentation provides a brief survey of the key ideas in this research field and is geared at both experimentalists and theorists. In particular, several theoretical mechanisms leading to deviations from relativity theory are presented; the standard theoretical framework for relativity violations at currently accessible energy scales (i.e., the SME) is reviewed, and various present and near-future experimental efforts within this field are discussed.
\end{abstract}


\maketitle

\section{Introduction}
\label{intro} 

Despite their phenomenological success, 
the Standard Model of particle 
physics and general relativity 
leave unresolved a number of theoretical questions. 
For this reason, 
a considerable amount of theoretical effort 
is currently directed toward the search for a more fundamental theory 
that includes a quantum description of the gravitational field. 
However, 
experimental tests of theoretical approaches to a quantum theory of gravity 
face a key issue of practical nature: 
most quantum-gravity effects in virtually all leading candidate models 
are expected to be minuscule 
as a result of Planck-scale suppression. 
For instance, 
measurements at presently attainable energies 
are likely to demand sensitivities 
at the $\sim10^{-17}$ level or better.
This mini course gives an overview of one recent approach to this issue 
involving the violation of spacetime symmetries. 

Due to the expected minute size of candidate quantum-gravity effects,
promising experimental avenues are difficult to identify. 
One idea in this context is testing physical laws 
that obey three key criteria. 
The first criterion is
that one should focus on fundamental laws
that hold {\it exactly} in established physical theories 
at the fundamental level.
Any observed deviations from these laws
would then definitely imply qualitatively new physics.
Second,
the likelihood of measuring such deviations is increased
by testing laws
that are predicted to be {\it violated}
in credible approaches to more fundamental physics.
The third criterion is a practical one:
for the potential to detect Planck-suppressed effects,
these laws should be amenable to {\it ultrahigh-precision} tests. 

One sample physics law
that satisfies all of these criteria
is CPT symmetry~\cite{cpt}.
As a brief reminder,
this fundamental law states 
that all physics remains invariant 
under the combined operations
of charge conjugation (C), 
parity inversion (P),
and time reversal (T).
Here, 
the C transformation links particles 
and antiparticles,
P denotes the spatial reflection 
of physics quantities
through the coordinate origin, 
and T reverses a given physical process in time. 
The Standard Model of particle physics 
is CPT symmetric by construction,
so that the first criterion is satisfied.
In the context of criterion two, 
we mention 
that a variety of candidate fundamental theories
can accomodate CPT violation. 
Such approaches include
string theory~\cite{kps},
spacetime foam~\cite{sf},
nontrivial spacetime topology~\cite{klink},
and cosmologically varying scalars~\cite{varscal}.
The third criterion above is met as well. 
Consider,
for example, 
the conventional figure of merit for CPT symmetry 
in the neutral-Kaon system:
its value lies presently at $10^{-18}$, 
as quoted by the Particle Data Group~\cite{pdg}.

Similar arguments can also be made 
for other spacetime symmetries, 
such as Lorentz and translational symmetry.
They are ingrained into our current understanding of physics
at the fundamental level;
they can be affected in various quantum-gravity approaches
because quantum gravity is likely
to require a radically different ``spacetime'' concept 
at the Planck length;
and being symmetries, 
ultrahigh-senstivity searches
for deviations from Lorentz and translational invariance
can be devised. 
The point is 
that tests of discrete and continuous spacetime symmetries 
have become a key tool in the phenomenology of new physics 
that possibly arises at the Planck scale.

This mini course is organized as follows. 
Section \ref{symmetries} 
discusses the relation between various spacetime symmetries. 
Two sample mechanisms for CPT- and Lorentz-symmetry violation 
in Lorentz-invariant underlying theories 
are reviewed in Sec.\ \ref{mechanisms}. 
The basic philosophy and ideas 
behind the construction of the Standard-Model Extension (SME) 
are presented in Sec.\ \ref{smesec}. 
Section \ref{tests} 
comments on a number of Lorentz and CPT tests 
in a variety of physical systems. 
A brief summary 
is contained in Sec.\ \ref{sum}.

\section{The interplay between different spacetime symmetries} 
\label{symmetries} 

Spacetime transformations can be divided into two distinct sets, 
namely discrete and continuous transformations.
The discrete transformations include C, P, and T
discussed in the introduction,
as well as various combinations of these, 
such as CP and CPT.
Examples of possible continuous transformations 
are translations, rotations, and boosts. 
If symmetry under one or more of these transformations is lost,
it is natural to ask 
as to whether the remaining transformations 
continue to be symmetries, 
or if the violation of one type of spacetime symmetry 
can lead to the breakdown of other spacetime invariances. 
This sections contains a few remarks about this issue. 

We begin by considering the CPT transformation. 
The renowned CPT theorem,
established by Bell, L\"uders, and Pauli, 
essentially states the following: 
under a few mild assumptions, 
symmetry under CPT 
is a consequence of quantum theory, locality, and Lorentz invariance. 
If deviations from CPT symmetry were observed in nature, 
one or more of the ingredients 
necessary to prove the CPT theorem 
must be incorrect. 
The question now becomes
which one of the key ingredients
that enter the CPT theorem 
should be dropped. 

The answer depends largely 
on the presumed underlying physics. 
But suppose the low-energy leading-order effects of new physics 
can be described 
within a local effective field theory.  
(Effective field theory is an enormously flexible tool. 
In the past,
it has been successfully applied in numerous contexts 
including condensed-matter systems, 
nuclear physics, 
and elementary-particle physics.)
It then seems unavoidable 
that exact Lorentz symmetry needs to be abandoned.
This expectation has recently been proven rigorously 
in the context of axiomatic quantum field theory
by Greenberg.
His ``anti-CPT theorem'' roughly states 
that in any unitary, local, relativistic point-particle field theory 
CPT violation comes with Lorentz breakdown \cite{green02,green}.
However, 
it is important to note
that the converse of this statement---namely 
that Lorentz violation implies CPT breakdown---does not hold true in general. 
In any case, 
we see 
that in the above general and plausible context,
CPT tests also probe Lorentz invariance.
We remark 
that other types of CPT violation 
arising from apparently non-unitary quantum mechanics
have also been considered~\cite{mav}.

We continue by supposing
that translational invariance is violated. 
This possibility can arise in the context 
of cosmologically varying scalar fields 
(see next section).
When translational symmetry is lost, 
the generator of translations
(i.e., the energy--momentum tensor $\theta^{\mu\nu}$)
is typically no longer a conserved current. 
We now turn to the question
as to whether Lorentz symmetry 
would be affected in such a scenario.
We begin 
by looking at the generator for Lorentz transformations, 
the angular-momentum tensor $J^{\mu\nu}$, 
which is given by 
\begin{equation} 
J^{\mu\nu}=\int d^3x \;\big(\theta^{0\mu}x^{\nu}-\theta^{0\nu}x^{\mu}\big)\;. 
\label{gen} 
\end{equation}
Note 
that this definition 
contains the energy--momentum tensor $\theta^{\mu\nu}$, 
which is not conserved 
in the present context.
As a result, 
$J^{\mu\nu}$ 
will generally exhibit a nontrivial dependence on time, 
so that the ordinary time-independent 
Lorentz-transformation generators no longer exist. 
For this reason, 
exact Lorentz symmetry 
is not guaranteed. 
It is apparent
that 
(with the exception of special cases) 
translation-symmetry breaking
leads to Lorentz-invariance violation.

\section{Sample mechanisms for spacetime-symmetry breaking} 
\label{mechanisms} 

In the previous section, 
we have argued 
that under certain circumstances the breakdown of one spacetime invariance
may lead to the violation of another spacetime symmetry.
Perhaps a more interesting question is 
how a translation-, Lorentz-, and CPT-invariant underlying model 
can lead to the breakdown of a particular spacetime symmetry 
in the first place.
The present section addresses this issue
by giving some intuition 
regarding mechanisms for spacetime-symmetry violation 
in candidate fundamental theories. 
Of the various possibilities for Lorentz breakdown 
mentioned in the introduction, 
we will focus on spontaneous Lorentz and CPT violation 
as well as Lorentz and CPT breaking through cosmologically varying scalars. 

\begin{figure}
\includegraphics[width=0.60\hsize]{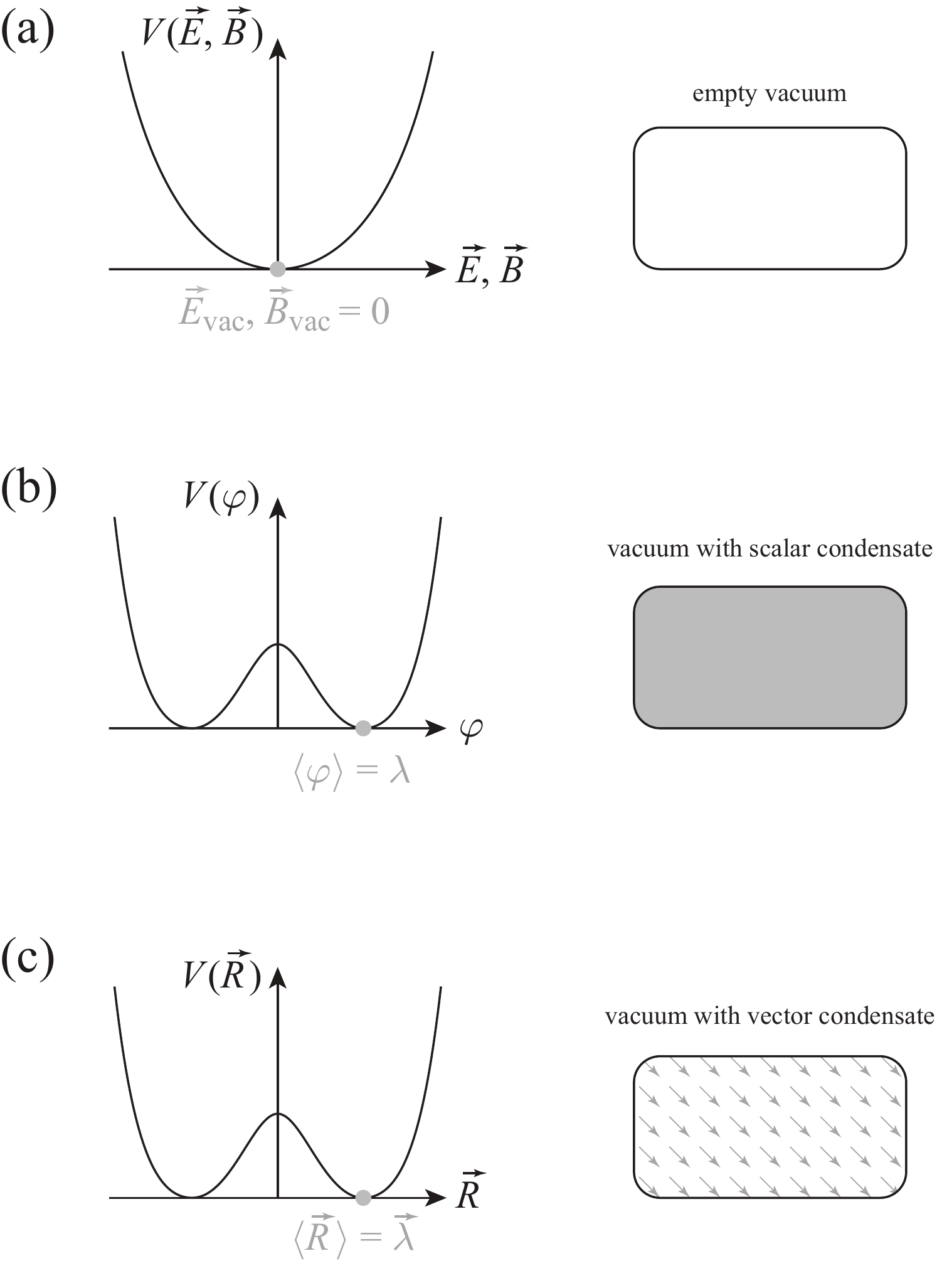}
\caption{Spontaneous symmetry breaking. 
In ordinary electrodynamics (a), 
the lowest-energy configuration is attained 
for zero $\vec{E}$ and $\vec{B}$ fields. 
The vacuum stays essentially free of fields. 
For a Higgs-type field (b), 
interactions generate an energy density $V(\varphi)$ 
that requires a non-vanishing value of $\varphi$ 
in the ground state. 
The vacuum contains 
a scalar condensate 
depicted in gray. 
Lorentz and CPT symmetry still hold
(but other, internal symmetries may be broken). 
Vector fields present in string theory (c), 
for example, 
can possess interactions similar to those of the Higgs 
demanding a nonzero field value in the lowest-energy state. 
The VEV of a vector field would select a preferred direction in the vacuum, 
which breaks Lorentz and possibly CPT invariance.}
\label{fig3} 
\end{figure}

{\bf Spontaneous Lorentz and CPT breakdown.} 
The mechanism of spontaneous symmetry breaking
is well established in many areas of physics
including
the physics of elastic media, 
condensed-matter physics, 
and elementary-particle theory.
From a theoretical perspective, 
this mechanism is very appealing
for the following reason.
Often, 
a symmetry is needed for the internal consistency of a quantum field theory,
but the symmetry is not observed in nature. 
This is exactly what spontaneous symmetry breaking achieves: 
At the dynamical level, 
the symmetry remains intact 
and ensures consistency.
It is only the ground-state solution 
(which pertains to experiments)
that is associated with the loss of the symmetry.
In order to gain intuition 
about the spontaneous breaking of Lorentz and CPT symmetry, 
we will look at three sample physical systems 
whose features will lead us step by step
to a better understanding of the effect. 
These three examples are 
illustrated in 
Fig.~\ref{fig3}. 

Let us first consider classical electrodynamics. 
Any electromagnetic-field configuration 
possesses an energy density $V(\vec{E},\vec{B})$, 
which is determined by
\begin{equation}
\label{max_en_den}
V(\vec{E},\vec{B})=\frac{1}{2} \left(\vec{E}^2+\vec{B}^2\right)\, .
\end{equation}
Here, 
natural units have been implemented, 
and $\vec{E}$ and $\vec{B}$ 
denote the electric and magnetic field, 
respectively. 
Equation (\ref{max_en_den}) 
yields the field energy 
of any given solution of the usual Maxwell equations. 
Notice 
that if the electric field, or the magnetic field, or both 
are different from zero in some region of spacetime, 
then the energy stored in these fields will be strictly positive. 
The field energy only vanishes 
when both $\vec{E}$ and $\vec{B}$ are zero throughout spacetime. 
The vacuum
is usually identified with the ground state, 
which is the lowest-energy configuration of a system. 
It is thus apparent 
that in conventional electrodynamics
the configuration with the lowest energy 
is the field-free one, 
so that the Maxwell vacuum is empty 
(disregarding Lorentz- and CPT-symmetric zero-point quantum fluctuations). 

Let us next consider a Higgs-type field. 
Such a field is contained in the phenomenologically very successful 
Standard Model of particle physics. 
As opposed to the electromagnetic field, 
which is a vector,
the Higgs field is a scalar. 
In our example, 
we will simplify various aspects of the model 
without distorting 
the features relevant in the present context. 
In the case for a constant Higgs-type field $\varphi$,
the expression for the energy density of $\varphi$
is given by 
\begin{equation} 
\label{scal_en_den} 
V(\varphi)=g (\varphi^2-\lambda^2)^2\, , 
\end{equation} 
where $\lambda$ and $g$ are constants. 
(A possible spacetime dependence  $\varphi=\varphi(x)$ 
would lead to additional, positive-valued contributions
to the energy density, 
so we can indeed focus on constant  $\varphi$.)
Paralleling the electrodynamics case described above, 
the lowest possible energy of $\varphi$ is zero. 
However, 
in contrast to the Maxwell example 
this lowest-energy configuration {\it requires} 
$\varphi$ to be nonzero: $\varphi=\pm\lambda$. 
As a result, 
the physical low-energy vacuum 
for a system involving a Higgs-type field $\varphi$
is not empty; 
it contains, in fact, 
the spacetime-constant scalar field 
$\varphi_{vac}\equiv\langle\varphi\rangle=\pm\lambda$.
Here, 
the quantity $\langle\varphi\rangle$ 
denotes the vacuum expectation value (VEV) 
of $\varphi$. 
We remark in passing
that a key physical effects 
of the VEV of the Standard-Model Higgs 
is to generate masses for many elementary particles. 
It is important to note
that $\langle\varphi\rangle$ is a scalar, 
and therefore it does {\it not} select a preferred direction in spacetime. 

Our final example is a 3-vector field $\vec{R}$. 
(The relativistic generalization to 4-vectors or 4-tensors is straightforward.)
This field (or its generalization)
is not contained in the Standard Model, 
and there is no observational evidence for such a field 
at the present time. 
However, 
additional vector fields like $\vec{R}$ 
are present in numerous candidate fundamental theories. 
Paralleling the Higgs case, 
we take the expression for the energy density of $\vec{R}=\textrm{const.}$
to be 
\begin{equation} 
\label{vec_en_den} 
V(\vec{R})=(\vec{R}^2-\lambda^2)^2\, . 
\end{equation} 
It is apparent 
that the lowest possible energy is zero, 
just as in the previous two examples
involving electromagnetism and a Higgs-type scalar.
As for the Higgs, 
this lowest-energy configuration necessitates a nonzero $\vec{R}$. 
In particular, 
we must require that $\vec{R}_{vac}\equiv\langle\vec{R}\rangle=\vec{\lambda}$, 
where $\vec{\lambda}$ is any constant vector obeying $\vec{\lambda}^2=\lambda^2$. 
As in the Higgs case, 
the vacuum does not stay empty; 
it rather contains the VEV of the vector field, $\langle\vec{R}\rangle$. 
Since we have only taken  
spacetime-independent solutions $\vec{R}$ into consideration, 
$\langle\vec{R}\rangle$ 
is also constant.
(A possible $x$ dependence would lead to 
positive definite derivative terms 
in Eq.\ (\ref{vec_en_den}) raising the energy density, 
as in the other two of the above examples.) 
Hence,
the true vacuum in the above model 
exhibits an intrinsic direction 
given by $\langle\vec{R}\rangle$. 
The upshot is 
that such an intrinsic direction
violates rotation invariance and thus Lorentz symmetry. 
We note 
that interactions generating energy densities like those in Eq.~\rf{vec_en_den} 
are absent in conventional renormalizable gauge theories, 
but they can be found in the context of string field theory, 
for instance. 

{\bf Spacetime-varying scalars.} 
A spacetime-dependent scalar, 
regardless of the mechanism causing this dependence, 
typically leads to the breakdown of spacetime-translation invariance \cite{varscal}.
In Sec.\ \ref{symmetries}, 
we have argued 
that translations and Lorentz transformations 
are closely linked in the Poincar\'e group, 
so that translation-symmetry violation 
typically leads to Lorentz breakdown. 
In the remainder of this section, 
we will focus on an explicit example for this effect. 

Consider a system with a spacetime-dependent coupling $\xi(x)$ 
and scalar fields $\phi$ and $\Phi$, 
and take the Lagrangian $\mathcal{L}$ 
to contain a kinetic-type term 
$\xi(x)\,\partial^{\mu}\phi\,\partial_{\mu}\Phi$. 
Under mild assumptions, 
one may integrate by parts the action for this system
(for instance with respect to the first partial derivative in the above term) 
without modifying the equations of motion. 
An equivalent Lagrangian $\mathcal{L}'$ would then be given by
\begin{equation}
\mathcal{L}'\supset -K^{\mu}\phi\,\partial_{\mu}\Phi\, ,
\label{example1}
\end{equation}
where $K^{\mu}(x)\equiv\partial^{\mu}\xi(x)$ is an external
prescribed 4-vector. 
This 4-vector clearly selects a preferred direction in spacetime 
breaking Lorentz invariance. 
We remark 
that for variations of $\xi(x)$ on cosmological scales, 
$K^{\mu}$ is spacetime constant locally (say on solar-system scales) 
to an excellent approximation. 

\begin{figure}
\includegraphics[width=0.50\hsize]{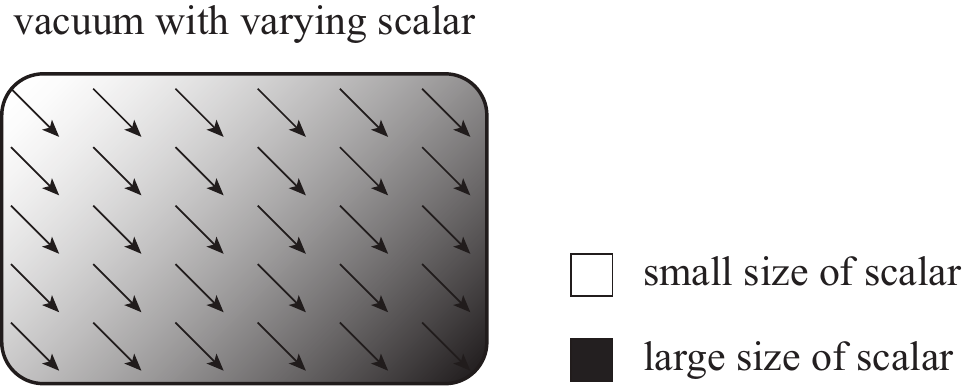}
\caption{Lorentz breakdown through spacetime-dependent scalars.
The background shading of gray measures the size of the scalar: 
the lighter regions correspond to smaller values of the scalar. 
The black arrows represent the gradient, 
which selects a preferred direction in the vacuum.
It follows that Lorentz symmetry is broken.}
\label{fig4} 
\end{figure}

The breakdown of Lorentz symmetry 
in the presence of a varying scalar can be understood intuitively as follows. 
The 4-gradient of the scalar has to be nonzero 
in some spacetime regions, 
for otherwise the scalar would be constant. 
This 4-gradient 
then singles out a preferred direction 
in such regions, 
as is illustrated in Fig.~\ref{fig4}. 
Consider, 
for example, 
a particle 
that possesses certain interactions with the scalar. 
Its propagation properties 
might be affected differently
in the directions perpendicular and parallel to the gradient. 
But physically inequivalent directions 
are associated with rotation-symmetry breaking. 
Since rotations are contained in the Lorentz group, 
Lorentz invariance must be violated. 

\section{The Standard-Model Extension}
\label{smesec}

In order to establish the low-energy phenomenology 
of Lorentz and CPT breaking 
and to identify 
relevant experimental signals for these effects, 
a suitable test framework is desirable. 
A number of Lorentz-symmetry tests are motivated and analyzed 
in purely kinematical models 
that describe small deviations from Lorentz invariance, 
such as Robertson's framework 
and its Mansouri--Sexl extension, 
the $c^2$ model, 
and phenomenologically constructed modified one-particle dispersion relations. 
However, 
the CPT properties of these test models lack clarity, 
and the absence of dynamical features 
greatly restricts their scope. 
To circumvent these issues, 
the SME, 
already mentioned in the introduction, 
has been developed. 
The present section contains a brief review 
of the philosophy behind the construction of the SME.

Let us first argue in favor of a dynamical rather than a purely kinematical test model. 
When the kinematical rules are fixed, 
there is certainly some residual freedom in introducing corresponding dynamical features.
However, 
the dynamics is constrained
by the requirement 
that established physics must be recovered 
in certain limits.
Moreover, 
it seems complicated 
and may not even be possible 
to construct an effective theory 
that contains the Standard Model 
with dynamics considerably different from that of the SME. 
We also mention 
that kinematical investigations
are limited 
to only a subset of potential Lorentz-violation signals 
emerging from fundamental physics. 
From this point of view, 
it appears to be desirable 
to implement explicitly dynamical features 
of sufficient generality 
into test frameworks for Lorentz and CPT invariance.  

{\bf The generality of the SME.}
In order to recognize 
the generality of the SME, 
we review the main ingredients of its construction~\cite{sme,AllOrders}.
Starting from the conventional Standard-Model and general-relativity Lagrangians 
${\mathcal L}_{\rm SM}$ and ${\mathcal L}_{\rm gr}$, respectively,
Lorentz-violating corrections $\delta {\mathcal L}$ are added: 
\begin{equation} 
{\mathcal L}_{\rm SME}={\mathcal L}_{\rm SM}+{\mathcal L}_{\rm gr}+\delta {\mathcal L}\; . 
\label{sme} 
\end{equation}
Here, 
${\mathcal L}_{\rm SME}$ denotes the SME Lagrangian. 
The correction terms $\delta {\mathcal L}$ 
are formed by contracting Standard-Model and gravitational fields  
of any mass dimensionality 
with Lorentz-breaking tensorial coefficients 
that describe a nontrivial vacuum 
with background vectors or tensors.
This background is presumed to originate 
from effects in the underlying theory,
such as those  discussed 
in the previous section. 
To ensure coordinate independence, 
these contractions must yield 
coordinate Lorentz scalars. 
We remark 
that in a curved-background context involving gravity, 
this procedure is most easily implemented 
employing the vierbein.
It thus becomes clear 
that all possible contributions to $\delta {\mathcal L}$ 
determine the most general effective dynamical description 
of first-order Lorentz violation 
at the level of observer Lorentz-invariant 
unitary effective field theory. 

Other potential features of underlying physics, 
such as non-pointlike elementary particles 
or a discrete spacetime structure at the Planck length, 
are  not likely to invalidate 
this effective-field-theory approach 
at presently attainable energies. 
On the contrary, 
the phenomenologically successful Standard Model 
and general relativity
are widely believed
to be effective-field-theory limits 
of more fundamental physics. 
If underlying physics 
indeed leads to minuscule Lorentz-breaking effects, 
it would appear somewhat artificial 
to consider low-energy effective models 
outside the framework of effective quantum field theory. 
We finally note 
that the requirement for a low-energy description 
beyond effective field theory 
is also unlikely to arise 
within the context of underlying physics 
with novel Lorentz-{\it symmetric} features, 
such as additional particles, 
new symmetries, 
or large extra dimensions. 
Note in particular 
that Lorentz-invariant modifications 
can therefore easily be implemented into the SME, 
should it become necessary~\cite{susy}. 

{\bf Advantages of the SME.}
The SME 
allows the identification 
and direct comparison 
of virtually all presently feasible experiments 
that search for deviations from Lorentz and CPT symmetry. 
Moreover, 
certain limits of the SME 
correspond to classical kinematics test models of relativity theory
(such as the previously mentioned framework by Robertson, 
its Mansouri--Sexl extension to arbitrary clock synchronizations, 
or the $c^2$ model)~\cite{km02,AllOrders}. 
A further benefit of the SME 
is the possibility of implementing 
additional desirable features 
besides coordinate independence. 
For example, 
one can choose to impose 
spacetime-translation invariance 
(at least in the flat-spacetime limit), 
SU(3)$\times$SU(2)$\times$U(1) gauge invariance, 
power-counting renormalizability, 
hermiticity,
and local interactions. 
These requirements 
place additional constraints on the parameter space for Lorentz and CPT breakdown. 
Another possibility is 
to make simplifying choices, 
such as a residual rotational symmetry 
in certain inertial frames. 
This latter hypothesis 
together with additional simplifications of the SME 
has been adopted in some investigations~\cite{cg99}.

\section{Lorentz and CPT tests}
\label{tests}

The full SME 
contains an infinite number of Lorentz- and CPT-violating coefficients.
However, 
in an effective field theory 
one might generically expect the power-counting renormalizable operators 
to dominate at low energies.
The restriction to this subset of the SME 
is called the minimal Standard-Model Extension (mSME).
To date, 
the flat-spacetime limit 
of the mSME
has been the basis 
for numerous 
phenomenological investigations 
of Lorentz and CPT violation
in many physical systems
including 
mesons \cite{kexpt,dexpt,bexpt,bexpt2,kpo,hadronth,ak},
baryons \cite{ccexpt,spaceexpt,cane},
electrons \cite{eexpt,eexpt2,eexpt3},
photons \cite{photon,km02}, 
muons \cite{muons}, 
and the Higgs sector~\cite{higgs}. 
Studies involving the curved-spacetime sector of the mSME 
have recently also been performed~\cite{grav}. 
We note 
that neutrino-oscillation measurements
harbor the potential for discovery~\cite{sme,neutrinos,nulong}.
This section contains a brief description 
of a representative sample of experimental efforts.

\begin{figure}
\includegraphics[width=0.75\hsize]{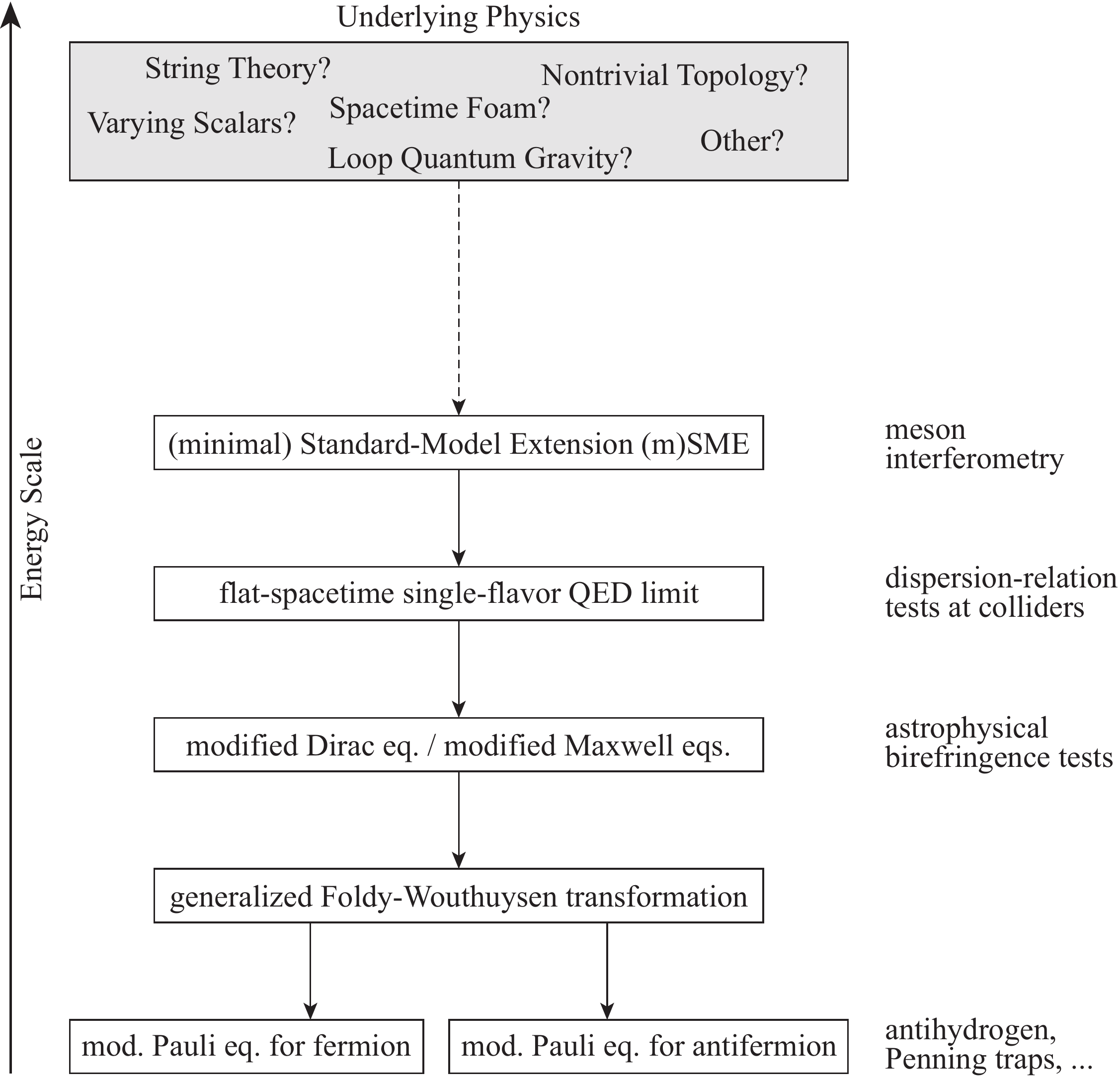}
\caption{Flow chart for phenomenological SME analyses. 
The neutral-meson effective Hamiltonian 
is derived from the quark sector of the mSME.
Collider tests involve photon decay, 
which requires a tree-level QFT calculation.
Spectropolarimetric studies of astrophysical sources 
are performed within the context of the classical (modified)
Maxwell equations.
CPT violation in the mSME
leads to different Pauli-type equations
for a fermion and its antifermion.
These modified Pauli equations 
are employed for calculations 
of atomic spectra, for example.}
\label{fig7} 
\end{figure}

{\bf Tests involving particle collisions.}
One of the key predictions of the mSME is 
that one-particle dispersion relations 
are typically modified by Lorentz and CPT violation. 
These modifications would result 
in changes to the kinematics of particle-collision processes.
(Energy--momentum remains conserved in the context of the flat-spacetime mSME 
because the Lorentz- and CPT-violating coefficients are taken as 
spacetime constant.)
For example, 
reaction thresholds may be shifted,
reactions kinematically forbidden in Lorentz-symmetric physics
may now occur,
and certain conventional reactions 
may no longer be allowed kinematically.

Consider, 
for example,
the spontaneous emission of a photon 
from a free electron.
In conventional physics, 
energy--momentum conservation 
does not allow this process to occur.
However, 
certain types of Lorentz and CPT breakdown
can slow down light relative to the speed of electrons.
In analogy to the usual Cherenkov effect
(when light travels slower inside a macroscopic medium with refractive index $n>1$),
electrons can then emit Cherenkov photons 
in such a Lorentz- and CPT-violating vacuum~\cite{cg99,cherenkov}.
This ``vacuum Cherenkov radiation'' 
may or may not be a threshold effect
depending on the type of mSME coefficient.
Let us consider mSME coefficients 
that are associated with a threshold 
for the vacuum Cherenkov effect.
In such a situation,
we obtain an observational constraint on the size of these mSME coefficients
as follows.
Electrons traveling with a speed 
above the modified light speed 
cannot do so for long: 
they would slow down below threshold 
through the emission of vacuum Cherenkov radiation. 
It follows 
that if highly energetic stable electrons exist in nature,
they must be below threshold. 
From this information 
one can extract a lower bound for the threshold,
which in turn gives a constraint on Lorentz breaking.
Employing LEP electrons with energies up to $104.5\,$GeV
in this context 
yields the limit $\ktr-\frac{3}{4}c^{00}\lsim1.2\times10^{-11}$~\cite{collider}.

Next, consider photon decay in vacuum.
This is another particle reaction process 
not allowed by energy--momentum conservation
in ordinary physics.
However, 
in the presence of certain mSME coefficients,
light may travel faster 
than the maximal attainable speed of electrons. 
In analogy to the above vacuum-Cherenkov case,
in which high-energy electrons become unstable, 
we expect 
that high-energy photons can now decay 
into an electron--positron pair.
With the modified dispersion relations 
predicted by the mSME,
one can indeed verify 
that this expectation is met.
As for vacuum Cherenkov radiation,
photon decay in a Lorentz-violating vacuum 
is often a threshold effect 
and can then be employed to extract 
an observational limit 
on this particular type of Lorentz breakdown. 
The idea is the following.
If stable photons are observed, 
they must essentially be below the decay threshold.
It then follows 
that the threshold energy must be higher 
than the energy of these stable photons.
This constraint on the threshold energy 
results in a limit on the size of the corresponding type of 
Lorentz violation.
At the Tevatron, 
stable photons with energies up to $300\,$GeV
were observed.
In this situation, 
our reasoning gives the bound 
$-5.8\times10^{-12}\lsim\ktr-\frac{3}{4}c^{00}$~\cite{collider}.

We note that the above results assume 
that both vacuum Cherenkov radiation and photon decay 
are efficient enough.  
The purely kinematical arguments 
we have presented 
are insufficient 
for conservative observational limits.
This is consistent with our remarks in the previous section 
that a dynamical framework is desirable,
and the full mSME 
(not only the predicted modified dispersion relations)
are needed.
Appropriate calculations within the mSME
indeed establish 
that the rates for vacuum Cherenkov radiation 
and photon decay 
would be fast enough 
to validate the above reasoning~\cite{ks08,collider}.

{\bf Spectropolarimetry of cosmological sources.}
The pure electrodynamics sector 
of the mSME 
contains one type of coefficient 
that violates both Lorentz and CPT invariance.
It is a mass dimension three term 
of Chern--Simons type 
parametrized by the $(k_{AF})^\mu$ background 4-vector. 
Among other effects, 
the $(k_{AF})$ term results in birefringence for photons~\cite{kAF}, 
the vacuum Cherenkov effect~\cite{cherenkov}, 
as well as shifts in cavity frequencies~\cite{cavity}. 
These deviations from established physics
are accessible to experimental investigations.
Birefringence searches in cosmic radiation are particularly well suited 
since the extremely long propagation time
is directly associated with an ultrahigh sensitivity 
to this type of Lorentz and CPT breakdown.
Spectropolarimetric studies of experimental data 
from cosmological sources 
have established a limit on $(k_{AF})^\mu$ 
at the level of $10^{-42}\,$GeV~\cite{kAF}.

{\bf Investigations of cold antihydrogen.} 
A comparison of the spectra of hydrogen (H) and antihydrogen ($\overline{\rm H}$) 
is well suited for Lorentz- and CPT-violation searches. 
There are a number of transitions 
that one can consider. 
One of them, 
the unmixed 1S--2S transition,  
seems to be an exquisite candidate: 
its projected experimental sensitivity is anticipated to be roughly 
at the level of $10^{-18}$, 
which is auspicious in light of the expected Planck-scale suppression 
of quantum-gravity effects. 
However, 
a leading-order calculation within the mSME
predicts identical shifts for free H or $\overline{\rm H}$ 
in the initial and final levels 
with respect to the conventional energy states. 
From this point of view, 
the 1S--2S transition is actually less satisfactory 
for the determination of unsuppressed Lorentz- and CPT-breaking effects.
Within the mSME,
the leading non-trivial contribution to this transition 
is generated by relativistic corrections, 
and it comes with two additional powers 
of the fine-structure constant $\alpha$. 
The predicted shift in the transition frequency, 
already expected to be minute at zeroth order in $\alpha$, 
is thus associated 
with a further suppression factor 
of more than ten thousand~\cite{antiH}. 

An additional transition 
that can be used for Lorentz and CPT tests 
is the spin-mixed 1S--2S transition.
When H or $\overline{\rm H}$ is trapped with electromagnetic fields
(e.g., in a Ioffe--Pritchard trap)
the 1S and the 2S states are each split 
as a result of the usual Zeeman effect. 
The mSME then predicts 
that in this case the 1S--2S transition 
between the spin-mixed states is indeed affected 
by Lorentz- and CPT-violating coefficients 
at leading order. 
A disadvantage from a practical perspective 
is the magentic-field dependence of this transition, 
so that the experimental sensitivity is limited 
by the size of the inhomogeneity of the $\vec{B}$ field in the trap. 
The development of new experimental techniques 
might avoid this issue, 
and a frequency resolutions close to the natural linewidth
could then be attained~\cite{antiH}. 

A third transition interesting from a Lorentz- and CPT-violation perspective
is the hyperfine Zeeman transitions within the 1S state itself.
Even in the limit of a zero $\vec{B}$ field, 
the mSME establishes leading-order level shifts 
for two of the transitions
between the Zeeman-split states.
We remark 
that this result may also be advantageous 
from an experimental perspective
because a number of other transitions of this type, 
such as the conventional H-maser line,
can be well resolved in the laboratory~\cite{antiH}. 

{\bf Tests in Penning traps.}
The mSME establishes not only 
that atomic energy levels can be affected 
by the presence of Lorentz and CPT breakdown, 
but also, for instance, 
the levels of protons and antiprotons 
inside a Penning trap. 
A perturbative calculation predicts 
that only one mSME coefficient 
(a CPT-violating $b^\mu$-type background vector, 
which is coupled to the chiral current of a fermion)
affects the transition-frequency shifts
between the proton and its antiparticle
at leading order.
To be more specific, 
the anomaly frequencies are displaced in opposite directions 
for protons and antiprotons. 
This effect can be used to 
determine a clean experimental constraint 
on the proton's $b^\mu$ coefficient~\cite{eexpt}.

{\bf Neutral-meson interferometry.} 
A long established standard CPT-symmetry test 
is the comparison of the K-meson's mass 
to that of the corresponding antimeson:
even extremely small mass differences 
would yield measurable effects in Kaon-interferometry experiments.
In spite of the fact that the mSME contains only one mass operator 
for a given quark species and the associated antiquark species, 
these (anti)particles are nevertheless influenced differently 
by the  Lorentz- and CPT-breaking background in the mSME. 
This causes the dispersion relations for a meson and its antimeson to differ, 
so that mesons and antimesons can possess distinct energies 
despite having equal 3-momenta. 
It is this split in energy 
that is ultimately measurable in interferometric experiments,
and it is thus potentially observable in such systems. 
We remark 
that not only the K-meson 
but also other neutral mesons can be investigated.
Note in particular 
that in addition to CPT breakdown, 
Lorentz violation is involved as well, 
so that boost- and rotation-dependent effects 
can be looked for~\cite{kexpt,dexpt,bexpt,bexpt2,kpo,hadronth,ak}.

\section{Summary}
\label{sum}

To date, 
no credible observational evidence 
for deviations from relativity theory exist.
However, 
in theoretical approaches to underlying physics, 
such as in models of the Planck-length structure of spacetime, 
minuscule violations of Lorentz and translation symmetry 
can be accommodated. 
In this mini course,  
we have given an overview of the
motivations, 
theoretical ideas, 
and experimental efforts 
in this spacetime-symmetry-breaking context.

We have argued 
that quantum-gravity models, 
for example, 
should describe a quantized version 
of the dynamics of spacetime.
In such a quantized spacetime, 
the concept of a smooth manifold 
may break down at some small distance scale,
so that the usual spacetime symmetries 
may only emerge at low energies.
We have reviewed  two specific examples
how Lorentz invariance might be violated: 
in string field theory 
(i.e., via spontaneous symmetry breaking)
or in the context of varying scalars
(i.e., via the gradient of the scalar).

At presently attainable energies 
and under mild assumptions regarding the dynamics,
the effects of general Lorentz and CPT breakdown
are described by an effective field theory 
called the SME. 
This framework contains essentially the entire body 
of established physics
(i.e., the Standard model and general relativity), 
so that predictions for Lorentz- and CPT-breaking effects
in essentially all physical systems are possible, 
at least in principle.
The coefficients for Lorentz and CPT violation in the SME 
are prescribed non-dynamical background vectors and tensors 
assumed to be generated by more fundamental physics.

Spacetime symmetries underpin numerous physical effects.
Accordingly, 
Lorentz and CPT invariance can be tested in a wide variety 
of physical systems. 
This fact, 
together with the generality of the SME
and the strong motivations for Lorentz and CPT violations, 
has led to a recent surge of experimental efforts 
to test relativity theory. 
We have reviewed 
a representative sample of these efforts 
in the contexts
of dispersion-relation studies, 
astrophysical polarimetry, 
and matter--antimatter comparisons.

A variety of important unanswered questions remain in this field.
They are of theoretical, of phenomenological, as well as of experimental nature, 
and they provide ample ground for further research in spacetime-symmetry physics.


\section*{Acknowledgments}
The author wishes to thank the organizers
for the invitation to present this mini course.
This work was funded in part 
by CONACyT under Grant No.\ 55310.


\end{document}